# The ODINUS Mission Concept

*The Scientific Case for a Mission to the Ice Giant Planets with Twin Spacecraft to Unveil the History of our Solar System*

**Spokeperson:** **Diego Turrini**
**Istituto di Astrofisica e Planetologia Spaziali INAF-IAPS**
**Via del Fosso del Cavaliere 100**
**00133 - Rome, Italy**
**Email: diego.turrini@iaps.inaf.it**

## *Authors list*

Diego Turrini[1], Romolo Politi[1], Roberto Peron[1], Davide Grassi[1], Christina Plainaki[1], Mauro Barbieri[2], David M. Lucchesi[1], Gianfranco Magni[1], Francesca Altieri[1], Valeria Cottini[3], Nicolas Gorius[4], Patrick Gaulme[5], François-Xavier Schmider[6], Alberto Adriani[1], Giuseppe Piccioni[1]

(1) Institute for Space Astrophysics and Planetology INAF-IAPS, Italy.
(2) Center of Studies and Activities for Space CISAS, University of Padova, Italy.
(3) University of Maryland, USA.
(4) Catholic University of America, USA
(5) New Mexico State University, USA
(6) Laboratoire Lagrange, Observatoire de la Côte d'Azur, France

## *Supporters list*

Roberto Orosei (INAF/IAPS, Italy), Giovanna Rinaldi (IAPS-INAF, Italy), Ernesto Palomba (IAPS-INAF, Italy), Alessandra Migliorini (IAPS-INAF, Italy), Maria Cristina De Sanctis (IAPS-INAF, Italy), Emiliano D'Aversa (INAF/IAPS, ITALY), Fabrizio Oliva (IAPS-INAF, Italy), Anna Milillo (INAF-IAPS, Italy), Agustín Sánchez-Lavega (Universidad del País Vasco UPV/EHU, Spain), Yves Langevin (Institut d'Astrophysique Spatiale, France), Patrick Irwin (University of Oxford, United Kingdom), Joern Helbert (DLR, Germany), J.-C. Gerard (LPAP- Univ. de Liege, Belgium), Rumi Nakamura (IWF/OEAW, Austria), Bortolino Saggin (Politecnico di Milano, Italy), Bianca Maria Dinelli (ISAC-CNR, Italy), Nikolay Ignatiev (Space Research Institute of Russian Academy of Sciences, Russia), David Luz (CAAUL/University of Lisbon, Portugal), Ioannis A. Daglis (Department of Physics - University of Athens, Greece), Glenn Orton (Jet Propulsion Laboratory- California Institute of Technology, USA), Alexander Rodin (Moscow Institute of Physics and Technology, Russia), Themelis Konstantinos (National Observatory of Athens, Greece), Britney E Schmidt (Georgia Institute of Technology, United States), Thomas B. McCord (Bear Fight Institute, USA), Randy Gladstone (SwRI, USA), H. MAVROMICHALAKI (University of Athens, Greece), anna maria fioretti (CNR - Institute of Geosceinces and Earth Resources, Italy), Kurt Retherford (Southwest Research Institute, USA), Michael Davis (Southwest Research Institute, United States of America), Bertrand Bonfond (Université de Liège, Belgium), Iannis Dandouras (IRAP, France), Dario De Fazio (IMIP-CNR, Italia), Leo Girardi (INAF - Osservatorio Astronomico di Padova, Italy), Tiziano Maestri (University of Bologna, Italy), Mauro Focardi (INAF-OAA, Italy), Jean-Baptiste Vincent (MPS, Germany), Neil Murphy (Jet Propulsion Lab, US), Mark Hofstadter (JPL/Caltech, United States), Anna Cinzia Marra (CNR-ISAC, ITALY), Vincenzo della Corte (Università degli Studi di Napoli Parthenope, Italy), Arnold (DLR, Germany), Gian Paolo Marra (CNR-ISAC , ITALY), Maurizio Pajola (CISAS - University of Padova, Italy), Maarten Roos (Lightcurve Films, Portugal), Cesare Grava (Southwest Research Institute, USA), and many others (see ODINUS website for the full list).

## *ODINUS website and full list of supporters*

http://odinus.iaps.inaf.it/

# Overview

The planets of our Solar System are divided in two main classes: the terrestrial planets, populating the inner Solar System, and the giant planets, which dominate the outer Solar System. The giant planets, in turn, can be divided between the gas giants Jupiter and Saturn, whose mass is mostly constituted by H and He, and the ice giants Uranus and Neptune, whose bulk composition is instead dominated by the combination of the astrophysical ices $H_2O$, $NH_3$ and $CH_4$ with metals and silicates. While in the case of the gas giants H and He constitutes more than 90% of their masses, in the case of the ice giants these gaseous envelopes are more limited, amounting to only 1-4 Earth masses (De Pater and Lissauer 2010). The terrestrial planets and the gas giants have been extensively studied with ground-based observations and with a large numbers of dedicated space missions. The bulk of the data on the ice giants, on the contrary, has been supplied by the Voyager 2 mission, which performed a fly-by of Uranus in 1986 followed by one of Neptune in 1989.

The giant planets appeared extremely early in the history of the Solar System, forming across the short time-span when the Sun was still surrounded by a circumstellar disk of gas and dust and therefore predating the terrestrial planets. The role of the giant planets in shaping the formation and evolution of the young Solar System was already recognized in the pioneering works by Oort and Safronov in 1950-1960. In particular, Safronov (1969) suggested that the formation of Jupiter would inject new material, in the form of planetesimals scattered by the gas giant, in the formation regions of Uranus and Neptune. More recently, the renewed understanding of planetary formation we obtained by the study of extrasolar planetary systems gave rise to the idea that the Solar System could have undergone a much more violent evolution than previously imagined (e.g. the Nice Model for the Late Heavy Bombardment, Tsiganis et al. 2005), in which the giant planets played the role of the main actors in shaping the current structure of the the Solar System.

The purpose of this document is to discuss the scientific case of a space mission to the ice giants Uranus and Neptune and their satellite systems and its relevance to advance our understanding of the ancient past of the Solar System and, more generally, of how planetary systems form and evolve. As a consequence, the leading theme of this proposal will be the first scientific theme of the Cosmic Vision 2015-2025 program:

- What are the conditions for planetary formation and the emergence of life?

In pursuing its goals, the present proposal will also address the second and third scientific theme of the Cosmic Vision 2015-2025 program, i.e.:

- How does the Solar System work?
- What are the fundamental physical laws of the Universe?

The mission concept we will illustrate in the following will be referred to through the acronym **ODINUS**, this acronym being derived from its main fields of scientific investigation: **Origins, Dynamics and Interiors of Neptunian and Uranian Systems**. As the name suggests, the ODINUS mission is based on the use of two twin spacecraft to perform the exploration of the ice giants and their regular and irregular satellites with the same set of instruments. This will allow to perform a comparative study of these two systems so similar and yet so different and to unveil their histories and that of the Solar System.

# Theme 1: What are the conditions for planetary formation and the emergence of life?

In this section we will briefly summarize how our understanding of the processes of planetary formation has evolved across the years, discuss their chronological sequence for what concerns the Solar System and highlight how the exploration of Uranus, Neptune and their satellite systems can provide deeper insight and better understanding of the history of the Solar System.

## The Evolving View of Planetary Formation: Solar System and Exoplanets

The original view of the set of events and mechanisms that characterize the process of planetary formation (Safronov 1969) was derived from the observation of the Solar System as it is today. This brought to the assumption that planetary formation was a local, orderly process that produced regular, well-spaced and, above all, stable planetary systems and orbital configurations. However, with the discovery of more and more planetary systems through ground-based and space-based observations, it is becoming apparent that planetary formation can result in a wide range of outcomes, most of them not necessarily consistent with the picture derived from the observations of the Solar System.

The orbital structure of the majority of the discovered planetary systems seems to be strongly affected by planetary migration due to the exchange of angular momentum with the circumstellar disks (see e.g. Papaloizou et al. 2007 and references therein), in which the forming planets are embedded, and by the so-called "Jumping Jupiters" mechanism (Weidenschilling & Marzari 1996; Marzari & Weidenschilling 2002), which invoke multiple planetary encounters, generally after the dispersal of the circumstellar disk, with a chaotic exchange of angular momentum between the different bodies involved.

The growing body of evidence that dynamical and collisional processes, often chaotic and violent, can dramatically influence the evolution of young planetary systems gave rise to the idea that also our Solar System could have undergone the same kind of evolution and represent a "lucky" case in which the end result was a stable and regular planetary system. The most successful attempt to describe the evolution of the Solar System to the present epoch has been the so-called Nice Model (Gomes et al. 2005; Tsiganis et al. 2005; Morbidelli et al. 2005; Morbidelli et al. 2007; Levison et al. 2011). The Nice Model is a Jumping Jupiter scenario formulated to link the event known as the Late Heavy Bombardment (LHB, see e.g. Hartmann et al. 2000 for a review) to a migration event involving all the giant planets.

In the Nice Model, the giant planets of the Solar System are postulated to be initially located on a more compact orbital configuration than their present one and to interact with a massive primordial trans-Neptunian region. The gravitational perturbations among the giant planets are initially mitigated by the trans-Neptunian disk, whose population in turn is eroded. Once the trans-Neptunian disk becomes unable to mitigate the effects of the interactions among the giant planets, the orbits of the latter become excited and a series of close encounters takes place. The net result of the Jumping Jupiters mechanism in the Nice Model is a small inward migration of Jupiter and marked outward migration of Saturn, Uranus and Neptune (Tsiganis et al. 2005).

The importance of the Nice Model lies in the fact that it strongly supports the idea that the giant planets did not form where we see them today or, in other words, that what we observe today is not necessarily a reflection of the Solar System as it was immediately after the end of its formation process. Particularly interesting in the context of this proposal is that, in about half the cases considered in the Nice Model scenario, Uranus and Neptune swapped their orbits (Tsiganis et al. 2005). The success of the Nice Model in explaining several features of the Solar System opened the road to more extreme scenarios, also based on the Jumping Jupiters mechanism, either postulating the existence of a now lost fifth giant planet (Nesvorny et al. 2011) or postulating an earlier phase of migration and chaotic evolution more violent and extreme than the one described in the Nice Model (Walsh et al. 2011).

In strict relation with the idea of the giant planets migration, one of the most fascinating aspects of these scenarios is that they all invoke a certain degree of mixing of the solid materials that compose the Solar System. The mixing is generally the larger the more the causing event is located toward the beginning of the Solar System lifetime. As an example, the "Grand Tack" scenario by Walsh et al. (2011, 2012) implies a much stronger remixing than the one that the LHB would cause in the framework of the Nice Model (see e.g. Levison 2009). However, a more or less extensive migration of the giant planets is not required to have a remixing of the solid material in the Solar Nebula. Safronov (1969) pointed out that the formation of Jupiter would scatter the planetesimals in its vicinity both inward and outward respect to its orbital region. The outward flux of ejected material was postulated to rise the density of solid material in the formation regions of Uranus and Neptune and increase their accretion rate.

The inward flux instead crosses the regions of the terrestrial planets and the asteroid belt, with potentially important implications for the collisional evolution of the primordial planetesimals (Weidenschilling 1975, Weidenschilling et al. 2001; Turrini et al. 2011, 2012). The influence of Jupiter's formation, however, is not limited to the scattering of neighboring planetesimals: the orbital resonances with the planet would extract planetesimals from farther away regions and put them on orbits crossing those of the other forming giant planets. One of the regions affected by the orbital resonances is the asteroid belt (Turrini et al. 2011, 2012): rocky material is therefore extracted from the inner Solar System and, as in the original idea from Safronov (1969), possibly accreted by the forming cores of Uranus and Neptune or captured in their circumplanetary disk and incorporated in their satellites.

## The Role of Ice Giants in Unveiling the Past of Solar System

As discussed in the previous section, during it history the Solar System went through a series of violent processes that shaped its present structure. The main actors of these processes were the giant planets. Due to their smaller masses and their likely later formation, Uranus and Neptune were also strongly affected by these very same processes. In this section, we will reorganize the events discussed in the previous section in a chronological order and discuss their implications for Uranus and Neptune and their satellite systems. If we follow the description of the history of the Solar System by Coradini et al. (2011), we can divide it into three main phases: the Solar Nebula, the Primordial and the Modern Solar System. This schematic view of the evolution of the Solar System is summarized in Fig. 1, where we reports the main events that took place across the different phases.

### The Solar Nebula

From the point of view of the giant planets, the Solar Nebula (see Fig. 1) is the period across which they were forming in the circumsolar disk and migrating due to disk-planet interactions. While the giant planets are forming, their gravitational perturbations on the protoplanetary disk cause a sequence of bombardment events that Coradini et al. (2011) called the Primordial Heavy Bombardment. One of the consequences of this Primordial Heavy Bombardment is that, after the formation of the first giant planet, each successive giant planet forms from a more and more evolved and remixed disk, in which the abundances of different elements and materials are different from the original ones, with implications for the rock/ice ratio and the ratio between different ices in the cores of the giant planets and in the material available for the forming satellites. In the standard view of the Solar System formation (Safronov 1969), the migration of the giant planets due to their exchange of angular momentum with the circumsolar disk was limited and the main role in reshuffling the protoplanetary disk was played by the Primordial Heavy Bombardment. However, in the alternative views we presented in the previous section, the migration of the giant planets could have played a significant role in the reshuffling of the different materials in the Solar System. In the "Grand Tack" scenario (Walsh et al. 2011, 2012) the giant planets are hypothesized to migrate extensively across the Solar System. Their formation regions, in this case, would be markedly different from those assumed by the standard scenario and the composition of their planetary cores would be affected by it. Moreover, part of the planetesimals that the giant planets scatter while migrating would collide with the giant planets themselves, contributing to the late accretion of high-Z elements first hypothesized by Owen et al. (1999) to explain the super-solar abundances of C, N, S, Ar, Kr and Xe in the atmosphere of Jupiter. All these remixing events, moreover, affect the source materials, captured in the form of planetesimals by the circumplanetary disks, from which the regular satellites of the giant planets can form (see Coradini et al. 2010 for a review). Depending on the formation time of the relevant giant planet and on the amount of radiogenic sources (incorporated in the rocky fraction of the source material), the regular satellites could already differentiate across this phase of the life of the Solar System. Finally, across the Solar Nebula phase a first generation of irregular satellites of the giant planets could have been captured from the protoplanetary disk due to collisions, the effects of gas drag or a combination of the two (see e.g. Mosqueira et al. 2010 for a discussion). This first generation of irregular satellites, however, would not survive the LHB if the latter is associated to a violent rearrangement of the Solar System like the one hypothesized by the Nice Model.

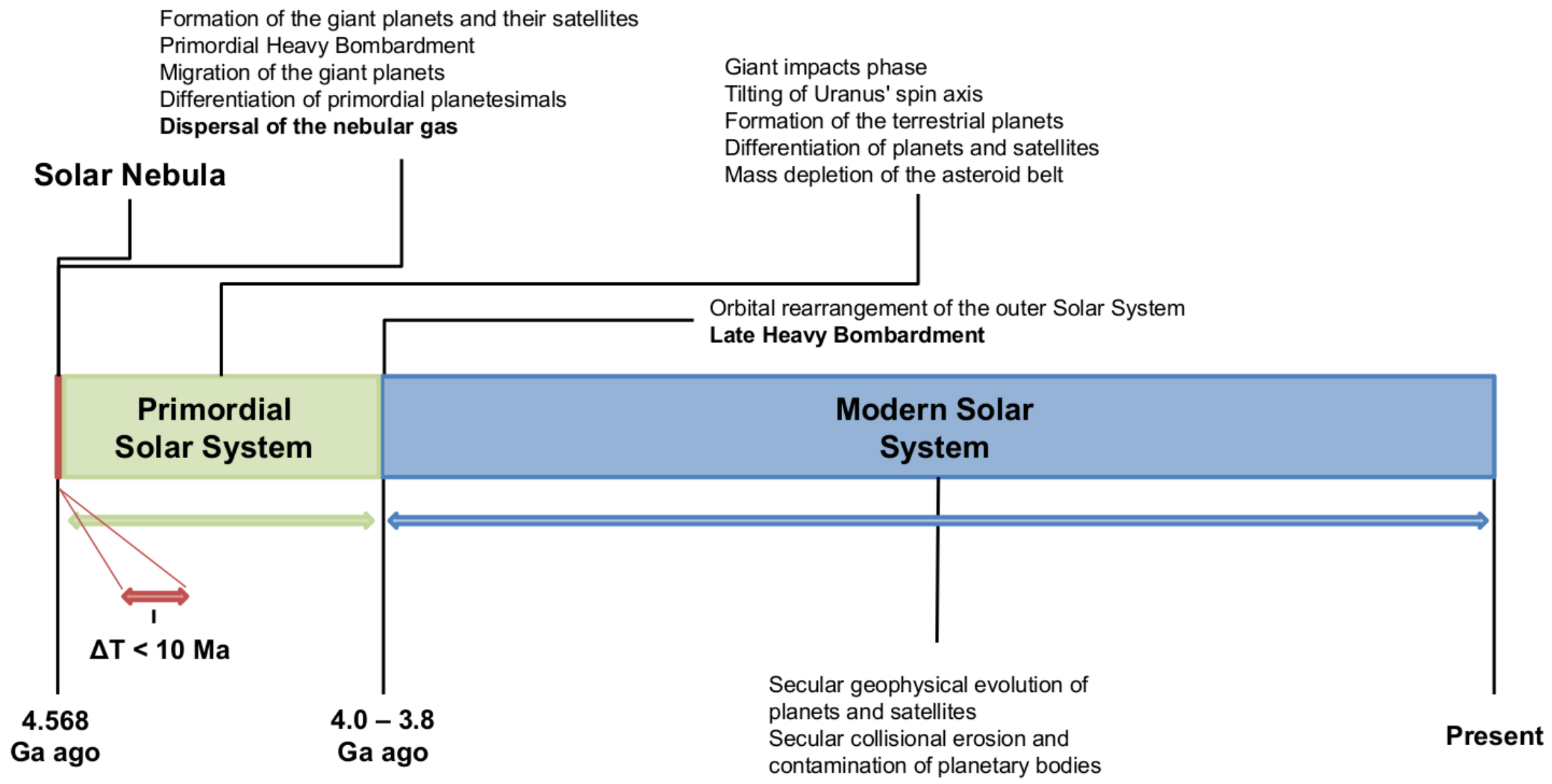

*Figure 1: time-line of the history of the Solar System following the division in three phases (Solar Nebula, Primordial and Modern Solar System) proposed by Coradini et al. (2011). The events marking the transition between the different phases are in bold characters.*

## The Primordial Solar System

Somewhere between the Solar Nebula and the Primordial Solar System phases, two events contributed to shape the Uranian and Neptunian satellite systems. One was the giant impact of a planetary embryo with Uranus, suggested to be responsible for its 98° obliquity. As discussed by Coradini et al. (2010), it is possible that the original satellite system of the ice giant was destroyed during this event and new satellites formed from the debris of the original ones. The second event was the capture of Triton by Neptune and the following shrinking and circularizing of its orbit, which caused the removal of most of the original regular satellites of the ice giant. Across these events and throughout the Primordial Solar System, the Nice Model predicts that the giant planets would still be on different, closer orbits with respect to their present ones. Once the dynamical instability responsible for the LHB takes place, icy planetesimals from what will become the trans-neptunian region are excited into high-eccentricity, giant planet-crossing orbits analogous to those of the present-day Centaurs. A fraction of these planetesimals will impact against the giant planets, possibly contributing to the late enrichment of their atmospheres (Matter et al. 2009). A fraction of these planetesimals will also impact on the satellites of the giant planets, contributing to their contamination by exogenous material and possibly supplying energy for their late differentiation (Barr & Canup 2010). Barr & Canup (2010) argue that the LHB can bring to the differentiation of Ganymede but not to that of Callisto, in agreement with the available data on their internal structure. Matter et al. (2009) assessed instead the amount of high-Z elements that would be accreted by the four giant planets during the LHB, finding that it is insufficient to explain or significantly contribute to the observed values. Another implication of the Nice Model is that any pre-existing population of irregular satellites would be destroyed as a consequence of the close encounters between the giant planets (Tsiganis et al. 2005). Nesvorny et al. (2007) however showed that three-body effects between the giant planets and the planetesimals during the planetary encounters invoked by the Nice Model would naturally supply a way to re-populate the satellite systems of the giant planets by irregular satellites. It must be noted that these studies are based on the earlier formulation of the Nice Model (Tsiganis et al. 2005; Gomes et al. 2005; Morbidelli et al. 2005) and that the implications of its more recent formulation (Morbidelli et al. 2007; Levison et al. 2011) are still to be addressed. Nevertheless, they show that the evolution of the Solar System across the Primordial Solar System phase could have a non-negligible role in shaping the present-day Uranus and Neptune and their satellite systems.

## The Modern Solar System

The Modern Solar System phase starts after the end of the LHB and, differently from the previous two phases, instead of violent processes it is dominated by more regular, secular ones. Moreover, the population of small bodies in the outer Solar System is significantly smaller than that at earlier times, so that collisional processes are less intense than before. Most of the information that we can gather through crater counting on the surface of the satellites of the giant planets refers to this long, more quiescent phase, especially if the satellites are still geophysically active and undergo resurfacing, as it appears to be the case of Triton (see Schubert et al. 2010 for a discussion). In the case of geophysically active satellites, moreover, the surface features and composition supply us information on their more recent internal state, i.e. they again give us insight on the processes that acted across the Modern Solar System phase. Depending on the degree of geophysical activity and the flux of impactors (being them planetocentric, i.e. other satellites, or heliocentric, e.g. comets and Centaurs), the surfaces of the satellites can be contaminated to various degrees by exogenous material (see e.g. Mosqueira et al. 2010, Schubert et al. 2010 for a discussion), an effect that has to be taken into account while interpreting e.g. spectral data, as spectrometers allow to probe the composition of a very thin layer (~cm-sized) of the satellites surfaces. Across the Modern Solar System, moreover, the secular effects of space-weathering due to various exogenic sources (e.g. solar wind, magnetospheric plasma, cosmic rays) contributed to the surface evolution of the satellites in ways that are still poorly quantified or even understood.

### The ODINUS mission and the history of the Solar System

As the previous sections highlight, our view of the processes of planetary formation and of the evolution of the Solar System has greatly changed across the last twenty years but most of the new ideas are in the process of growing to full maturity or need new observational data to test them against. The ODINUS mission aims to address these open problems by exploring the systems of Uranus and Neptune, as they are the most affected from the violent processes that sculpted the early Solar System and yet they are the least explored and more mysterious ones.

The primary information that the ODINUS mission wants to gather by exploring the Uranian and Neptunian systems are:

- What is the atmospheric composition and enrichment with respect to solar abundances of the two planets?
- What are the bulk densities and the masses of the ice giants and their satellites?
- What are the interior structures and density profiles of the ices giants and their satellites?
- What is the surface composition of the regular and irregular satellites?
- Which satellites are fully or partially differentiated and which ones are undifferentiated?

Using these data, the open questions that ODINUS aims to answer are:

- When and where did the planets form? Did they migrate? If so, how much? Did Uranus and Neptune swap their positions as hypothesized by the Nice Model?
- What is the ice-to-rock ratio of the cores of ice giants and of their satellites? How much “non-local” material was available to them when they formed? Where did this “non-local” material originated from?
- Are the satellites of Uranus primordial or they reformed after the planet tilted its spin axis? What were the effects of the capture of Triton for the Neptunian satellites?
- Where did the irregular satellites originate? Can they be used to constrain the dynamical evolution of the ice giants?

## Theme 2: How does the Solar System work?

In gathering the data that will allow to address Theme 1 of the Cosmic Vision 2015-2025 program, the ODINUS mission will gather a wealth of data on the present status of the Uranian and Neptunian systems. While the twin spacecraft setup constrains the number of instruments on-board each spacecraft, the goal of the ODINUS mission is to perform a global survey as complete as possible of the two giant planets and their satellites. The data, which ODINUS will collect, will allow to gain a more complete understanding of how icy satellites so far away from the Sun evolve both for what it concerns their surfaces and their interiors. Moreover, the coupled investigation of these two planets, so similar and yet so different, will allow to better understand the sources of their different atmospheric and thermal behavior.

## Atmospheres of Uranus and Neptune

The Herschel observations of Uranus and Neptune (Feuchtgruber et al., 2013) confirmed that the ice giants have a remarkably similar D/H content ($4.4 \pm 0.4 \times 10^{-5}$ and $4.1 \pm 0.4 \times 10^{-5}$ respectively), suggesting a common source of icy planetesimals in the protoplanetary disk. Further insight on the conditions of the disk in its outer regions can be derived from the relative enrichment (with respect to the Solar values) of C, N, S and O, by determination of the abundances of the corresponding reduced forms. At the current date, methane is still the only minor atmospheric constituent that has been directly detected in both ice giants (e.g.: Baines et al., 1994); an extensive investigation in this field is therefore extremely urgent to ultimately characterize the emergence of our solar system.

The post-Voyager 2 observations of Uranus by ground-based and space telescopes revealed a progressive increase of meteorological activity (cloud and dark spots occurrence) in the proximity of Northern Spring equinox (see, e.g. Sromovsky et al., 2012). While this evolution is undoubtedly related to the extreme obliquity of the planet, the relative roles of solar illumination and internal heating (and its possible variations) remain to be assessed by detailed studies at high spatial resolution.

The possibility to compare the atmospheric behaviour of Uranus with the extremely dynamic meteorology of Neptune – apparently characterized by a slower long-term evolution – provides a unique opportunity to gain insights on the response of thick atmospheres to time-variable forcing, representing therefore a new area of tests for future atmospheric global circulation models, in conditions not found in terrestrial planets or gas giants.

Uranus zonal winds are currently characterized by moderately retrograde values (-50 m/s) at the equator that progressively become prograde, to reach a maximum value of 200 m/s at 50N (Sromovsky et al., 2012). On Neptune, a similar pattern is observed, but the absolute speed values are strongly amplified, to reach – despite the limited solar energy input – the extreme values (400m m/s or more) observed in the Solar System (Shuleen Chau et al., 2012). Wind speed fields are the most immediate proxy for atmospheric circulation and their modeling can provide constraints on very general properties of the atmosphere, such as the extent of deep convection (Suomi et al., 1991).

While the efforts of ground based observers has allowed to considerably expand the results of Voyager 2, an extensive, long-term, and high spatial resolution cloud tracking remains essential to study the ultimate causes of these extreme phenomena.

Neptune shows an unexpected temperature of 750 K in its stratosphere (Broadfoot et al., 1989) that cannot be justified by the small solar UV flux available at that heliocentric distance. More complex mechanisms – such as energy exchange with magnetospheric ions – shall become predominant in these regions. Uranus, on the other hand, offers unique magnetospheric geometries because of its high obliquity and strong inclination of magnetic axis.

## The satellites of Uranus and Neptune

The satellites of Uranus and Neptune are poorly known, mostly due to the limited coverage and resolution of the Voyager 2 observations. The Uranian satellites Ariel and Miranda showed a complex surface geology, dominated by extensional tectonic structures linked to their thermal and internal evolution (Prockter et al. 2010 and references therein). Umbriel appeared featureless and dark, but the analysis of the images suggests an ancient tectonic system (Prockter et al. 2010 and references therein). Little is known about Titania and Oberon, as the resolution of the images taken by Voyager 2 was not enough to distinguish tectonic features. The partial coverage of the surface of Triton revealed one of the youngest surfaces of the Solar System, suggesting the satellite is possibly more active than Europa (Schubert et al. 2010 and references therein). Notwithstanding this, the surface of Triton showed a variety of cryovolcanic, tectonic and atmospheric features and processes (Prockter et al. 2010 and references therein).

From the point of view of their surface composition, the Uranian satellites are characterized by the presence of crystalline $H_2O$ ice (Dalton et al. 2010). The spectral features of Ariel, Umbriel and Titania showed also the presence of $CO_2$ ice, while $CO_2$ ice was not observed on Oberon (Dalton et al. 2010 and references therein). In the case of Miranda, the possible presence of ammonia hydrate was observed but both the presence of the spectral band and its interpretation are to be confirmed (Dalton et al. 2010 and references therein). The confirmation of the presence of ammonia would be of great importance due to its anti-freezing role in the interior of the satellites. The spectra of Triton possess the absorption bands of five ices: $N_2$, $CH_4$, CO, $CO_2$, and $H_2O$ (Dalton et al. 2010). The detection of the HCN ice band has been reported, which could imply the presence of more complex materials of astrobiological interest (see Dalton et al. 2010 and references therein). Triton also possesses a tenuous atmosphere mainly composed by $N_2$ and CO, which undergoes seasonal cycles of sublimation and recondensation (see Dalton et al. 2010 and references therein). Images taken by Voyager 2 revealed active geyser-like vents on the surface of Triton, indicating that the satellite is still geologically active even if at present it is not tidally heated (Schubert et al. 2010).

Both Uranus and Neptune possess a family of irregular satellites. Neptune, in particular, possesses the largest irregular satellite in the outer Solar System (not counting Triton), i.e. Nereid. Aside their estimated sizes and the fact that observational data suggest they might be more abundant than those of Jupiter and Saturn (Haghighipour and Jewitt 2007), almost nothing is known of these bodies.

## Magnetosphere-Exosphere-Ionosphere Coupling in the Uranian and Neptunian systems

The highly non-symmetric internal magnetic fields of Uranus and Neptune, coupled with the relatively fast rotation and the unusual inclination of the rotation axes to the orbital planes imply that their magnetospheres are subject to drastic geometrical variations on both diurnal and seasonal timescales. The relative orientations of the planets' spin axis, their magnetic dipole axis and the direction of the solar wind flow determine the configuration of each magnetosphere and, consequently, the plasma dynamics in these regions.

Due to the planet's large obliquity, Uranus' asymmetric magnetosphere varies from a pole-on to orthogonal configuration during an Uranian year (84 Earth years) and changes from an "open" to a "closed" configuration during an Uranian day. At solstice (when Uranus' magnetic dipole simply rotates around the vector of the direction of the solar wind flow) plasma motions produced by the rotation of the planet and by the solar wind are effectively decoupled (Selesnick and Richardson, 1986; Vasyliunas, 1986). Moreover, the Voyager 2 plasma observations showed that when the Uranus dipole field is oppositely directed to the interplanetary field, injection events to the inner magnetosphere (likely driven by reconnection every planetary rotation period) are present (Sittler et al., 1987). The time-dependent modulation of the magnetic reconnection sites, the details of the solar wind plasma entry in the inner magnetosphere of Uranus and the properties of the plasma precipitation to the planet's exosphere and ionosphere are unknown. Models indicate that Uranus' ionosphere is dominated by $H^+$ at higher altitudes and $H_3^+$ lower down (Capone et al., 1977; Chandler and Waite, 1986; Majeed et al., 2004), produced by either energetic particle precipitation or solar ultraviolet (UV) radiation. Our current knowledge on the aurora of Uranus is limited since it is based only on: one spatially resolved observation of the UV aurora (by the Ultraviolet Spectrograph data on board Voyager 2, Herbert 2009); observations of the FUV and IR aurora with the Hubble Space Telescope (Ballester, 1998); and on observations from ground-based telescopes (e.g., Trafton et al., 1999). The details of the solar wind plasma interaction with the planet's exosphere, ionosphere and upper atmosphere (through charge-exchange, atmospheric sputtering, pick-up by the local field), the seasonal and diurnal variation of the efficiency of each mechanism as well as the total energy balance (deposition/loss) due to magnetosphere-exosphere-ionosphere coupling are unknown. Since the exact mechanism providing the required additional heating of the upper atmosphere of Uranus is also unknown, new in situ plasma and energetic neutral particles observations could become of particular importance in order to determine whether plasma precipitation play a key role in this context. The magnetospheric interaction with the Uranian moons can be studied through in situ measurements of magnetic field, particles, and energetic neutrals emitted from the surfaces. Finally, remote imaging of charge exchange energetic neutral atoms would offer a unique opportunity to monitor the plasma circulation where moons and/or Uranus exosphere are present.

Neptune's magnetic field has a complex geometry that includes relatively large contributions from non-dipolar components, including a strong quadrupole moment that may exceed the dipole moment in strength. Unlike Uranus, however, Neptune has shown no evidence of UV emission that could be associated with auroral activity. Although this non-observation did not rule out an active magnetosphere *per se*, it ruled out processes similar to those associated with the aurora observed at Uranus. Whereas the plasma in the magnetosphere of Uranus has a relatively low density and is thought to be primarily of solar-wind origin, at Neptune, the distribution of plasma is generally interpreted as indicating that Triton is a major source (Krimigis et al., 1989; Mauk et al., 1991, 1994; Belcher et al., 1989; Richardson et al., 1991). Escape of neutral hydrogen and nitrogen from Triton maintains a large neutral cloud (Triton torus) that is believed to be source of neutral hydrogen and nitrogen (Decker and Cheng, 1994). The escape of neutrals from Triton could be an additional plasma source for the Neptune's magnetosphere (through ionization). Our knowledge on the plasma dynamics in the magnetosphere of Neptune as well as on the neutral particles production in Triton's atmosphere is limited. New in situ plasma and energetic neutral particles observations focused in the Triton region can provide important information on the role of the combined effects of photoionization, electron impact ionization, and charge exchange in the context of the coupling of a complex asymmetric planetary magnetosphere with a moon exosphere at large distances from the Sun.

## Planetary and satellite interiors

The available constraints on interior models of Uranus and Neptune are limited. The gravitational harmonics of these planets have been measured only up to fourth degree (J2, J4), and the planetary shapes and rotation periods are not well known (see e.g. Helled et al. 2011 and references therein). The response coefficients of Uranus and Neptune suggest that the latter is less centrally condensed than the former (De Pater and Lissauer 2010).

The thermal structures of these planets are also intriguing (see e.g. Helled et al. 2011 and references therein). Uranus stands among the planets for the extremely low value of 0.042 ± 0.047 W/m$^2$ of its internal energy flux (Pearl et al., 1990). This figure sharply contrasts with Neptune, where Voyager 2 determined a value of 0.433± 0.046 W/m$^2$ (Pearl et al., 1991). The two ice giants must therefore differ in their internal structure, heat transport mechanisms, and/or in their formation history. Substantial differences in internal structures are suggested by the analysis of available gravitational data for the two planets (Podolak et al., 1995).

Namely, the Uranus gravity data are compatible with layered convection in the shell, which inhibits the transport of heat. Alternative views call – among the others – for a later formation age of Neptune (Gudkova et al., 1988). Consequently, heat fluxes represent, along with gravity and magnetic data, the key experimental constraints to characterize the interior of Uranus and Neptune and their evolution.

The information on the interior structure of the satellites of Uranus and Neptune is even more limited and is mostly derived from their average densities, which are used to infer the rock-to-ice ratios, and their surface geology, which suggests that across their lives they possessed partially or completly molten interiors (De Pater and Lissauer 2010). As a consequence, the data that can be collected by the ODINUS mission on their interiors will play an important role in filling up this gap in our understanding of the icy satellites in the outer Solar System.

Gravity data can indeed be used to constrain the internal structure and composition of the planets. Deviations of the primary body gravitational field from the spherical symmetry (due to its rotational state and internal structure and composition) perturb the orbit of the spacecraft and can be extracted via a precise orbit determination and parameter estimation procedure from the tracking data, usually the range and the range rate in a typical Radio Science Experiment. Fundamental to this objective is a proper modeling of the spacecraft dynamics, both gravitational (e.g., gravitational multipoles) and non gravitational (e.g., radiation pressure). This could be non trivial in case of a complex spacecraft (the ideal would be a test mass) and – in selected cases – could require also the use of an on-board accelerometer (Iafolla et al., 2010). In the case of Uranus measurements of the precession of its elliptical rings should add to the list of observables. What said for the primaries extends to their satellites as well. Selected fly-bys to the satellites will allow for the determination of at least their lowest-degree multipoles.

An alternate and complementary method to probe the internal structures of Uranus and Neptune consists of using seismic techniques that were developed for the Sun (helioseismology, see e.g. Goldreich & Keeley 1977), then successfully applied to stars with the CoRoT and Kepler space missions (Michel et al. 2008, Borucki 2009), and tested on Jupiter (Gaulme et al. 2011). Seismology consists of identifying the acoustic eigen-modes, whose frequency distribution reflects the inner sound speed profile. The main advantage of seismic methods with respect to gravity moments is that waves propagate down to the central region of the planet, while gravitational moments are mainly sensitive to the external 20% of the planetary radius. The second advantage is that the inversion problem is not model dependent, neither on the equation of state or on the abundances that we want to measure. As regards Uranus and Neptune, the difference in internal energy flux should appear as a difference in the amplitude of acoustic modes. As for helioseismology, two approaches may be used to perform such seismic measurements, either with Doppler spectro-imaging (e.g. Schmider et al. 2007), or visible photometry (Gaulme & Mosser 2005). A dedicated study must be led to determine which method is the most appropriate for these two planets.

## Heliosphere science

During the ODINUS mission cruise phase, it will be possible to obtain important information on the interplanetary medium properties at different distances from the Sun as well as on the heliosphere structure and its interactions with the interstellar medium. Although there is plenty of information on how solar wind and coronal mass ejections interact with the interplanetary medium at 1 AU from the Sun, little is known on how this interaction works at larger distances. The ODINUS measurements of the interplanetary magnetic field fluctuations and plasma densities variations, at different distances from the Sun, can provide information for understanding the origin of turbulence in the solar wind and its evolution from its source to the heliopause. ODINUS, therefore, will give an opportunity to study space weather in the outer heliosphere and to understand how the interplanetary medium properties are modified in space and time.

The prevailing models of the shape of the heliosphere suggest a cometary-type interaction with a possible bow shock and/or heliopause, heliosheath, and termination shock (Axford, 1973; Fichtner et al., 2000). However, recent energetic neutral atom images obtained by the Ion and Neutral Camera (INCA) onboard Cassini did not conform to these models (Krimigis et al., 2009). Specifically, the map obtained by Cassini/INCA revealed a broad belt of energetic protons with non-thermal pressure comparable to that of the local interstellar magnetic field (Krimigis et al., 2009). In October 2008, Interstellar Boundary Explorer (IBEX) was launched with energetic neutral atom cameras specifically designed to map the heliospheric boundary at lower (<6 keV) energies (McComas et al., 2009; Funsten et al., 2009). Both IBEX and INCA identified in the energetic neutral atom images dominant topological features (ribbon or belt) that can be explained on the basis of a model that considers an energetic neutral atom-inferred non-thermal proton pressure filling the heliosheath from the termination shock to the heliopause (Krimigis et al., 2009).

During the cruise phase, the two spacecraft can be used measure the energetic neutral atoms produced by energetic singly charged particles in the heliosheath that charge-exchange with interstellar neutral hydrogen and enter the heliosphere unimpeded by the interplanetary magnetic field (Hsieh et al., 1992; Gruntman et al., 2001). Using also magnetic field measurements, the ODINUS can address the question whether the interaction of the heliosphere with the interstellar magnetic field takes place at the termination shock or at the heliopause.

### How well do we know the distribution of mass in the Kuiper Belt?

The cruise phase of the two spacecraft to Uranus and Neptune offers the possibility to improve our current knowledge of the total mass and the mass distribution of the Kuiper Belt. Among the various methods used for constraining this distribution, the study of heliocentric orbits of objects in the Solar System (Anderson et al., 1995) applies well to ODINUS. The spacecraft approaching Uranus and Neptune in their cruise could be considered (as in the fundamental physics experiments) as test masses subject to the gravitational attraction of the Kuiper belt objects: the accurate tracking of the spacecraft will therefore help to further constrain the total mass and the mass distribution of these objects.

## Theme 3: What are the fundamental physical laws of the Universe?

Since the early interplanetary exploration missions, spacecraft are used as (nearly) test masses to probe the gravitational machinery of Solar System and more in general to test for fundamental physics. Though general relativity is currently regarded as a very effective description of gravitational phenomena and it has passed all the experimental tests (both in the weak- and strong-field regimes) so far, it is challenged by theoretical (e.g. Grand Unification, Strings) scenarios and by cosmological findings (Turyshev, 2008). Stringent tests of general relativity have been obtained in the past by studying the motion of spacecraft in cruise, as well as the propagation of electromagnetic waves between spacecraft and Earth (see e.g. Bertotti et al. 2003). In this respect, the spacecraft are considered as test mass subject (mainly) to the gravitational attraction of Solar System bodies. Well-established equations of motions can then be tested against the experimental data, in order to place strong constraints to possible deviations from what is predicted by general relativity. Also for what it concerns electromagnetic waves propagation experiments, the spacecraft act as a virtual bouncing point for microwave pulses, enabling a measure of the Shapiro time delay. Being very effective in the past in ruling out possibilities of "exotic physics" (i.e., the so-called "Pioneer Anomaly"), such tests could be used in the future to further pursue experiments in this way. The very-weak-field environment of the more external regions of the Solar System is particularly interesting, in that “exotic” phenomenology such as MOND could be probed. These tests would help extend the scale at which precision information on gravitational dynamics is available; this will contribute to bridge the “local” scale (in which precise measurements on gravitational dynamics are available) to more “global” scales (subject to puzzling phenomenology as dark matter and dark energy).

## Scientific rationale of the twin spacecraft approach

The approach proposed for the ODINUS mission is to use a set of twin spacecraft (see Fig. 2), each to be placed in orbit around one of the two ice giant planets. The traditional approach for the exploration of the giant planets in the Solar System is to focus either on the study of a planetary body and its satellites (e.g. the Galileo and Cassini missions to the Jovian and Saturnian systems) or on the investigation of more specific problems (e.g. the Juno mission to study the interior of Jupiter and the JUICE mission to explore the Jovian moons Ganymede, Callisto and Europa). This is a well tested approach that allows for a thorough investigation of the subject under study and to collect large quantities of highly detailed data. The only drawback of this approach is that comparative studies of the different giant planets are possible only after decades, especially since the datasets provided by the different missions are not necessarily homogeneous or characterized by the same level of completeness, as the different missions generally focus on different investigations. In the case of the well-studied Jovian and Saturnian systems, about 10 years passed before it became possible to compare the dataset supplied by the Galileo mission with the first data supplied by the Cassini mission.

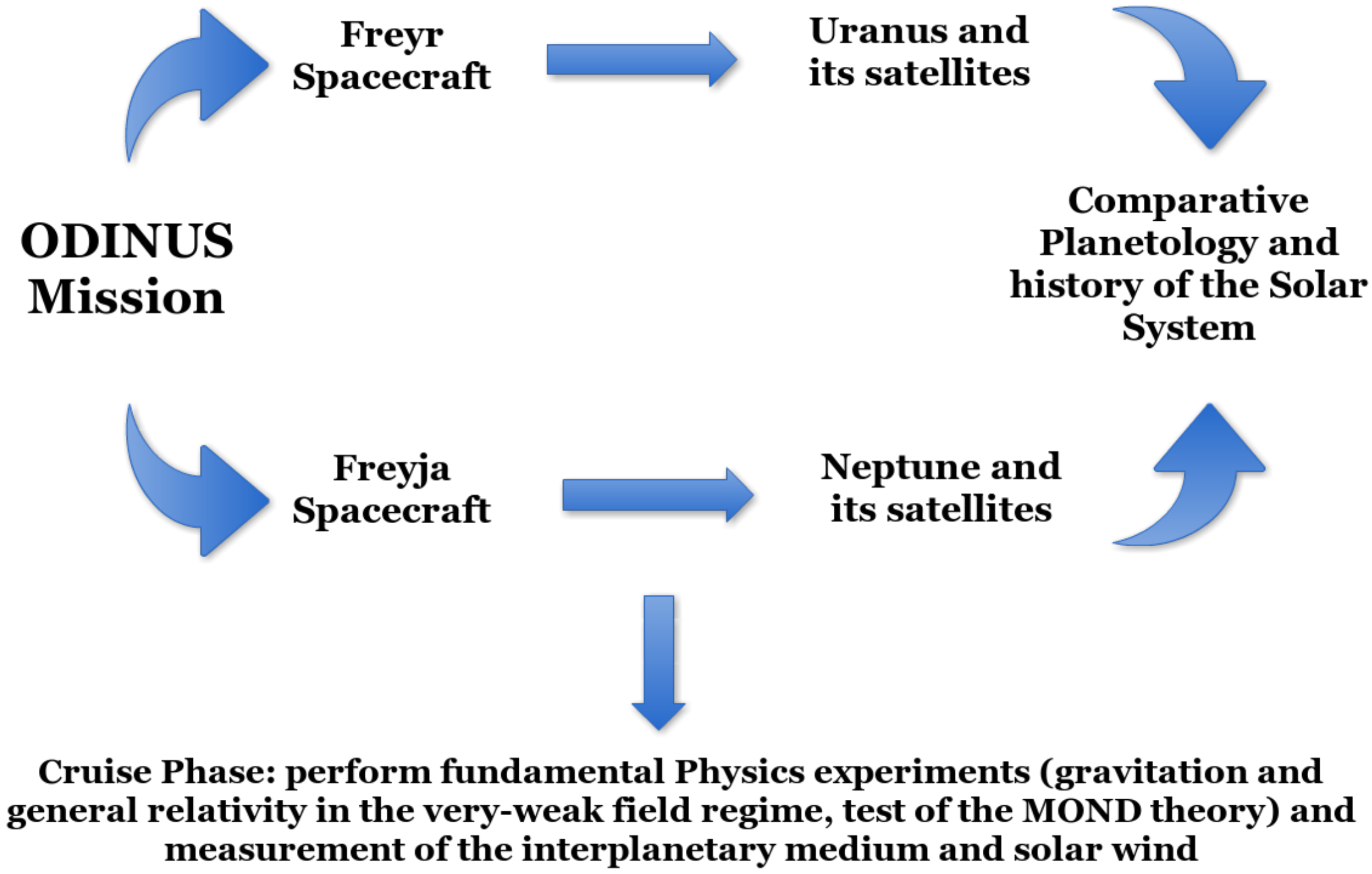

*Figure 2: schematics of the twin spacecraft approach of the ODINUS mission.*

However, in order to be able to perform a detailed comparative study of the satellites of these two giant planets it will be necessary to wait until the completion of the JUICE mission, due to the limited coverage of the data from Galileo. As a consequence, about half a century will be required before we can fully address the differences and similarities between the Jovian and Saturnian systems.

Exploring the Uranian and Neptunian systems with the traditional approach would require either half a century of efforts or the focus on this exclusive goal over the L2 and L3 missions of Cosmic Vision. In a scenario in which, to balance between the different needs of the astrophysics community, ESA would devote the L3 and L5 missions to the exploration of these two giant planets, the launch of the L5 missions would occur in 2046 or later (assuming a temporal distance between L5 and L4 and between L4 and L3 analogous to that between L3 and L2): assuming a travel time to Uranus and Neptune of about 13-15 years, as in the scenarios assumed for the Uranus Pathfinder (Arridge et al. 2012) and OSS (Christophe et al. 2012) mission proposals, the completion of the two missions would occur in about 2060, i.e. about half a century from now. In the unrealistic scenario of devoting both L2 and L3 missions to the exploration of the ice giants, it would be possible to complete this task by about 2050 but at the cost of not having L-class missions devoted to astrophysics before L4.

The approach proposed for the ODINUS mission is different from the traditional one in that it focuses on the use of two M-class spacecraft to be launched toward two different targets in the framework of the same mission. The use of two twin spacecraft, aside limiting the development cost of the mission, will allow to perform measurements with the same set of instruments in the Uranian and Neptunian systems, supplying data of similar quality and potentially completeness. Obviously, the need to produce and manage two spacecraft in place of one will limit the amount of instruments to be included in the scientific payload: this will translate in a less in-depth exploration of the two systems with respect to what would be possible with two dedicated missions. As we will discuss in the template mission profile, a careful selection of the instruments and design of the spacecraft can limit the importance of this drawback. Finally, we want to emphasize that, due to the different travel time to reach the two planets, the two spacecraft will not be operating at the same time except for short periods during the cruise phase, thus limiting the complexity of the mission management.

## A model mission profile for ODINUS

To illustrate the feasibility and the critical aspects of the ODINUS mission concept, in the following we will discuss a model mission profile. We will illustrate the possible configuration of the two spacecraft and their scientific payload, the orbital paths that we believe could maximise the scientific return and the launch slot that could best suit the ODINUS mission.

### The twin spacecraft

As we mentioned previously, the founding idea of the ODINUS mission concept is to have a set of twin spacecraft (which we dubbed Freyr and Freyja from the twin gods of the Norse pantheon) to be placed in orbit of Uranus and Neptune respectively. In order to fit the budget of an L-class mission, a conservative, straw-man configuration for the ODINUS mission could be based on two New Horizons-like spacecraft, i.e.:

- about 6 instruments in the scientific payload + radio science;
- about 500-600 kg of dry mass for each spacecraft;
- hybrid (ionic and chemical) propulsion;
- radioisotope-powered energy source.

The limitations on the scientific payload and the dry mass of the spacecraft come from a worst-case scenario evaluation of the fuel budget needed to reach the ice giants and to insert them on planetocentric orbits. If we consider the Hohmann transfer orbit between Earth and Uranus (or Neptune) with an orbital insertion at about $2x10^7$ km from the relevant planet on a highly eccentric orbit, the required $\Delta v$ of about 5 km/s translates into a wet-to-dry mass ratio of about 5 for each spacecraft. This implies that 600 kg of dry mass requires a wet mass at launch of about 3000 kg. Such a wet mass at launch would make the mission feasible either considering a single launch of the Freyr and Freyja spacecraft with an Ariane V rocket or two separate launches with Soyuz rockets. The scenario contemplating two separate launches allows the two trajectories to be optimized independently, thus allowing for the largest savings of either fuel or travel time, but a preliminary check of the orbital positions of Uranus and Neptune showed that the two ice giants will be in a favorable position to launch the two spacecraft together and then separate their paths at Uranus.

## The post-insertion orbital paths of the spacecraft and the exploration strategy of the Uranian and Neptunian systems

The choices of the insertion orbit and of the hybrid propulsion system are motivated by the exploration strategy of the Uranian and Neptunian systems. The basic idea is to have the spacecraft enter their planetocentric orbits thanks to the chemical propulsion and then to take advantage of the ionic propulsion to slowly spiral inward toward the respective planets. The insertion orbits are chosen to insert the spacecraft in the orbital regions populated by the irregular satellites and have one or more fly-bys with members of this family of small bodies. The spacecraft will then spiral toward the regions populated by the regular satellites, possibly maintaining highly eccentric orbits to allow for the contemporary observation of the regular satellites and the planets or their ring systems.

The high obliquity values of Uranus and Neptune imply that the regular satellites orbit on planes significantly inclined with respect to the ecliptic plane. As a consequence, unless the fuel budget and the orbital studies indicate the possibility of inserting the spacecraft on high-inclination orbits, the orbital path of the spacecraft will need to be optimized to allow for as many close encounters as possible with the regular satellites in the lifetime of the mission. This is particularly important in the case of Uranus, where the satellites orbit almost perpendicularly to the ecliptic plane: a spacecraft orbiting near the latter would therefore allow only for short close encounters with the regular satellites when they are approaching and crossing the ecliptic plane itself.

A possible solution could be to take advantage of the ionic propulsion to make the orbits of the spacecraft precess: the resulting rosetta orbit should be optimized to allow the most close-encounters with the regular satellites. After the completion of the exploration of the regular satellites, the spacecraft would shrink their orbits again in order to approach the planets and focus the next phase of the mission to their study. A possible end-mission scenario would then be to take advantage of the ionic propulsion to slowly spiral the spacecraft inside the atmospheres of the planets and use the two spacecraft as atmospherics probes. If feasible, the use of the ionic propulsion to slow down the atmospheric descent would allow to circumvent the needs of heath shields on the spacecraft, thus reducing their weight.

## The straw-man payload

A possible straw-man payload for the two spacecraft, which could allow for the achievement of the goals of the ODINUS mission, is composed by:

- Camera (Wide and Narrow Angle);

- VIS-NIR Image Spectometer;
- Magnetometer;
- Mass Spectrometer (Ions and Neutrals, INMS);
- Doppler Spectro-Imager (for seismic measurements) or Microwave Radiometer;
- Radio-science package.

The choice to limit the number of instruments on-board the spacecraft is due to the budget constraint, i.e. to the need of keeping the ODINUS mission inside the total cost for an L-class mission. As we will discuss also in the next section, given the long times required to explore the ice giant planets, the development of a highly integrated payload, in order to maximize the number of instruments that can be fit in the spacecraft and thus the scientific return of the mission, is critical for the success of ODINUS. Two instruments that would significantly improve the completeness of the exploration of Uranus and Neptune and their satellites and the scientific return of the mission would be:

- Energetic Neutral Atoms Detector (to complement the measurements of the INMS);
- High-sensitivity Accelerometer (for the atmospheric descent phase).

As discussed in the section devoted to the study of the planetary interiors, an alternative approach based on seismologic measurements can be coupled to the more traditional study of the gravitational momenta to study the interiors of Uranus and Neptune. The ODINUS mission would be the ideal test-bed for this new kind of measurements, as the launch slot we suggest (2034, as discussed in the next section) would allow to assess which of the possible approaches (doppler-spectro imaging or visible photometry) is the most appropriate for ODINUS. Should visible photometry prove to be the technique of choice, the Doppler-Spectro Imager we indicated in the straw-man payload could be replaced by one (or more) of the alternative instruments we discussed (microwave radiometer, ENA detector, accelerometer).

## Launch slot and timeline of the ODINUS mission

Given the technological challenges that the two-spacecraft approach of the ODINUS mission rises and the need to assess how to include seismological measurements among those performed by the spacecraft, we think that the optimal slot for ODINUS would be as the L3 mission of ESA Cosmic Vision 2015-2025 program, with the indicative launch foreseen for 2034. This would allow for enough time to develop the required enabling technologies (e.g. the radioisotope-powered energy source or a flight-qualified doppler spectro imager) and nevertheless, assuming an indicative time of flight of about 9 years to reach Uranus and 12 years to reach Neptune as achieved by the Voyager 2 mission, to complete the exploration of the outer Solar System by the first half of the century.

## Critical aspects and enabling technologies of the ODINUS mission

As we highlighted in the previous sections, the ODINUS mission is in principle feasible with the present-day technology. The two spacecraft are modeled after the one of the ongoing New Horizons mission and their wet masses, according to our first order estimates, would fit either the Soyuz (two launches scenario) or the Ariane V (single launch scenario) payload capabilities. With an estimated final cost of about 550 MEuro (source: NASA) for the New Horizons mission and taking into account that the development costs would be shared between the two spacecraft, the ODINUS mission would be feasible also from the point of view of the expected cost.

The two most critical aspects for the success of the ODINUS mission are:

- the availability of radioisotope-powered energy sources;

- the possibility to achieve times of flight comparable with those of the Voyager 2 mission.

The first critical aspect is due to the large distances of Uranus and Neptune from the Sun, which make the use of solar panels for energy generation unpractical: the development of the required technology and the identification of an affordable and reliable energy source compliant with ESA's policies is therefore mandatory for the feasibility of the ODINUS mission. The second aspect is not critical for the feasibility of the mission: the Uranus Pathfinder (Arridge et al. 2012) and OSS (Christophe et al. 2012) mission studies already showed that the mission could be feasible even if on longer timescales (13-15 years of time of flight). Nevertheless, the duration of the mission is of major importance since it determines the possibility to perform a comparative study of the two systems in a reasonable timespan as well as it influences the management cost of the mission.